\documentclass[11pt,a4paper]{article}
\pdfoutput=1
\usepackage{empheq}
\usepackage{cases}
\usepackage{a4wide}
\usepackage{amsfonts}
\usepackage{amssymb}
\usepackage{amsmath,bm}
\usepackage{amsmath}
\usepackage{graphicx}
\usepackage{mathtools}
\usepackage{booktabs}
\usepackage{array}
\usepackage{color}
\usepackage{ulem}
\usepackage[small,bf]{caption}
\usepackage{cite}
\usepackage[usenames,dvipsnames]{xcolor}
\usepackage{subfigure}
\usepackage{verbatim}

\allowdisplaybreaks

\numberwithin{equation}{section}

\newcounter{mysubequation}[equation]

\DeclarePairedDelimiter\ket{\lvert}{\rangle}
\DeclarePairedDelimiterX\braket[2]{\langle}{\rangle}{#1 \delimsize\vert #2}

\begin{document}
\begin{titlepage}

\begin{center}
{
\bf\LARGE Millicharge Dark Matter Detection\\[0.3em]
with Mach-Zehnder Interferometer
}
\\[8mm]
Chuan-Ren~Chen\footnote[1]{crchen@ntnu.edu.tw},~Bui~Hong~Nhung\footnote[2]{nhungqft@gmail.com},~and Chrisna~Setyo~Nugroho\footnote[3]{setyo13nugros@ntnu.edu.tw}  
\\[1mm]
\end{center}
\vspace*{0.50cm}

\centerline{\it Department of Physics, National Taiwan Normal University, Taipei 116, Taiwan}
\vspace*{1.20cm}

\begin{abstract}
\noindent
If the dark sector exists and communicates with Standard Model through the $U(1)$ mixing, it is possible that electromagnetism would have influence on matter fields in dark sector, so-called millicharge particles (mCPs). Furthermore, the highest mCPs could be dark matter particles.  Recently it has been shown that the mCPs would be slowed down and captured by the earth. As a result, the number density of accumulated mCPs underground is enhanced by several orders of magnitude as compared to that of dark matter in our solar system. In this study, we propose to use the Mach-Zehnder (MZ) laser interferometer to detect earth bound mCPs through the detection of phase shifts of photons. We show that, for mass of mCPs lager than $1$ GeV, the sensitivity of probing the mixing parameter $\epsilon$ could reach as low as $10^{-11}$ if number density is larger than $1~\rm{cm^{-3}}$.
\end{abstract}

\end{titlepage}
\setcounter{footnote}{0}

\section{Introduction}

One of the undisputed departure from the standard model (SM) of particle
physics is the existence of enigmatic matter alias dark matter (DM). 
Based on several observations, ranging from galactic rotational
curve to gravitaional lensing, one infers that it interacts
gravitationally with ordinary matter. Thus, it is widely believed
that DM should be electromagnetically neutral or chargeless. On the
theoretical side, however, the unobserved magnetic monopole implies
the violation of charge quantization pointing toward new particles  
with non-quantized charges. Moreover, there exists several studies
which assert that these millicharged particles (mCPs) are viable DM
candidates that
account for some or all of the observed DM abundance despite of its
minuscule charge~\cite{Dvorkin:2019zdi,Creque-Sarbinowski:2019mcm}.

There have been a number of  attempts to
explore the existence of mCPs. In laboratory frontier, the SLAC
millicharge experiment~\cite{Prinz:1998ua}, neutrino experiment~\cite{Magill:2018tbb}, BEBC beam dump
experiment~\cite{Marocco:2020dqu}, miliQan pathfinder experiment at the LHC~\cite{Ball:2020dnx} as well as the
Argoneut experiment~\cite{ArgoNeuT:2019ckq} has placed stringent constraints on mCPs mass in
MeV to TeV regime. On the other hand, the null results from
anomalous emission in stellar environments put strong limits on
mCPs mass in less than MeV range~\cite{Davidson:2000hf,Chang:2018rso}. In addition, there are several novel strategies to probe mCPs~\cite{Knapen:2017ekk,Blanco:2019lrf,Essig:2019kfe,Berlin:2019uco,Kurinsky:2019pgb,SENSEI:2020dpa,Griffin:2020lgd}. Furthermore, mCPs also escape the
conventional direct detection experiments such as XENON1T~\cite{XENON:2020rca}. 

Due to to its electromagnetic (EM) interaction with ordinary matter,  
the mCPs could have a large transfer cross section leading to the loss
of its virial kinetic energy and further thermalize with the environment.
As a result, when mCPs arrive at the detector of direct search experiment, it deposits insufficent energy to the detector.
Since these mCPs have lost its energy, they are trapped thanks to
the earth's gravity. This mechanism  leads to the terrestrial mCPs
accumulation taking place during the earth's existence. Furthermore,
for mCPs with masses larger than 1 GeV, they are sunk to the earth's
core leading to significant mCPs number density underground~\cite{Neufeld:2018slx,Pospelov:2020ktu}. 

Since mCPs interact with the photon, 
laser interferometer experiments offer a suitable venue for terrestrial mCPs
search. When mCPs interact with the laser in one arm of the
interferometer, it would induce a phase shift on the laser to be detected at the
output port. However, the existing DM search proposals utilizing the laser interferometer employed at Gravitational Wave (GW) experiments~\cite{Tsuchida:2019hhc,Lee:2020dcd,Chen:2021apc,Ismail:2022ukp,Lee:2022tsw,Seto:2004zu,Adams:2004pk,Riedel:2012ur,PhysRevLett.114.161301,Arvanitaki:2015iga,Stadnik:2015xbn,Branca:2016rez,Riedel:2016acj,Hall:2016usm,Jung:2017flg,Pierce:2018xmy,Morisaki:2018htj,Grote:2019uvn} are not
suitable for this. In typical interferometer at GW experiment, both
of the interferometer arms are located at the same depth
underground. Consequently, there is the same amount of mCPs in both
arms leading to zero phase shift in the photon path.  
We propose a phase measurement scheme based on optical laser
experiment using Mach-Zehnder (MZ) interferometer to explore these
mCPs. One arm of the MZ interferometer is located underground while keeping another arm on the earth surface.  We demonstrate that our proposal is more  sensitive
than the current cosntraints on heavy mCPs mass regime given by
collider experiments and the recent ion trap proposal~\cite{Budker:2021quh}.

The  paper is organized as follows: In section~\ref{sec:ebdm}, we
give a brief discussion of  terrestrial mCPs accumulation. We
proceed to examine how mCPs interact with photon in Section~\ref{sec:interaction}. We introduce a phase measurement based on MZ
interferometer in Section~\ref{sec:phase} and further present the
projected sensitivities of our proposal 
in section~\ref{sec:result}.
Our summary  are presented in Section~\ref{sec:Summary}.

\section{A Brief Review of Earth Bound millicharge Dark Matter}
\label{sec:ebdm}

The introduction of mCPs can be naturally realized in various ways.
One possibility is through the mixing between a $U(1)'$ Abelian
gauge field in dark sector and the SM hypercharge field. As the
matter fields in dark sector are charged under the $U(1)'$\cite{Okun:1982xi}, 
these particles would have interactions with the
photon with a coupling strength proportional to the mixing parameter
and their charges under $U(1)'$.
Phenomenologically, one can parametrize the electromagnetic coupling
of mCPs to be $\epsilon e$, where $e$ is the electric charge of
electron.  As a charged particle passes through the earth, it will
be slowed down due to the scattering with the ordinary matters
through its EM interaction, and even be stopped inside the
earth. Recently, Ref.~\cite{Pospelov:2020ktu} shows that, if it
constitutes partial or total amount of dark matter, the number density of mCPs could be several orders of magnitude higher than
that of the dark matter around our solar system. We summarize the
relevant conclusions to our study in this section, and refer readers
to~\cite{Pospelov:2020ktu} for more details and other cases.  

For the mass of mCPs we are interested in, namely $m_Q\gtrsim 1$ GeV, the average number density of mCPs on the earth is given as 
\begin{align}
\label{eq:avedensity}
<n_Q> \simeq <n_Q^{cap}> \approx f_Q\left(\frac{t_\oplus}{10^{10}~\rm year}\right)\left(\frac{\rm GeV}{m_Q}\right)\left(\frac{3\times 10^{15}}{\rm cm^3}\right),
\end{align}  
where $f_Q$ is the fraction of mCPs in total local DM density and $t_\oplus$ is the age of earth, since the evaporation can be
neglected. For number density underground, one needs to estimate the
transfer cross section $\sigma_T$ between mCPs and terrestrial
medium. Moreover, due to the gravitational pulling, mCPs would reach
terminal velocity been estimated as 
\begin{align}
\label{eq:vterm}
v_{\rm term} &= \frac{3m_Q g T}{m^2_{\rm rock}n_{\rm rock}<\sigma_T v^3_{\rm th}>}~~{\rm for}~~m_Q>m_{\rm rock} \\
{} &= \frac{m_Q g}{3 n_{\rm rock}T}\left<\frac{v_{\rm th}}{\sigma_T}\right>~~{\rm for}~~m_Q<m_{\rm rock},
\end{align}  
where $m_{\rm rock}$ and $n_{\rm rock}$ are the mass and number density of terrestrial medium atom, respectively, and  $v_{\rm th}$ is the thermal velocity of mCPs after thermalizing with atmosphere. Since the terminal velocity is slower than $v_{\rm vir}$ that is the average velocity of galactic mCPs, a so-called traffic jam effect causes an enhanced number density $n_{\rm tj}$ given by 
\begin{equation}
\label{eq:ntj}
n_{\rm tj} = \frac{v_{\rm vir}}{v_{\rm term}}n_{vir}
\end{equation}
where $n_{\rm vir}$ is the number density of galactic mCPs. Finally, the number density of mCPs underground can be estimated as 
\begin{equation}
\label{eq:ntj}
n_{\rm loc} = {\rm Max} \left(n_{Jeans}, {\rm Min}(n_{\rm tj}, \left<n_Q\right>)\right)\,,
\end{equation}
where $n_{Jeans}$ refers to the number density governed by the Jean's equation for a static, steady-state distribution of mCPs~\cite{Neufeld:2018slx}. 

Following~\cite{Budker:2021quh}, we assume that all mCPs considered
here are
free of binding and spreading everywhere. 
We focus on the case of heavy mCPs ($m_Q\gtrsim 1$ GeV) and take
benchmark values for accumulated number density $n_{\rm loc}=$ $1~{\rm cm^{-3}}$, $10^3~{\rm cm^{-3}}$, and $10^6~{\rm cm^{-3}}$ in our numerical study.
 
\section{mCPs and Photon Interaction}
\label{sec:interaction}

To probe the millicharge dark matter in a laser experiment, we start with  the Hamiltonian describing the
interaction between non-relativistic charged particles and the photon 
\begin{align}
\label{eq:hamiltontot}
H = H_{P} + H_{R} + H_{I} \,.
\end{align}
Here $H_{P}$, $H_{R}$, and $H_{I}$ denotes the free charged
particles, the free radiation field, and the interaction between charged particles and the radiation,
respectively. They are given by~\cite{cohen:1987}
\begin{align}
\label{eq:hamilall}
H_{P} &= \sum_{\alpha} \frac{\vec{p}^{2}_{\alpha}}{2\, m_{\alpha}} + V_{\text{Coulomb}}\\
H_{R} &= \sum_{i} \hbar \omega_{i} \left( \hat{a}^{\dagger}_{i} \hat{a}_{i} + \frac{1}{2}\right)\\
H_{I} &= H_{I1} + H_{I2}\\
H_{I1} &= - \sum_{\alpha} \frac{\text{q}_{\alpha}}{m_{\alpha}} \, \vec{p}_{\alpha} \cdot \vec{A}(\vec{r}_{\alpha})\\
H_{I2} &= \sum_{\alpha} \frac{\text{q}^{2}_{\alpha}}{2\,m_{\alpha}} \left[ \vec{A}(\vec{r}_{\alpha})\right]^{2}\,.
\end{align}  
where $\vec{p}_{\alpha}$, $m_{\alpha}$, and $\text{q}_{\alpha}$
stand for the momentum, the mass, and the electric charge of the  $\alpha$-th charged particle, respectively. The operator $\hat{a}_{i}$ ($\hat{a}^{\dagger}_{i}$) describes the annihilation (creation)
operator of the photon field for the i-th mode that satisfies the
commutation relation $[\hat{a}_{i},\hat{a}^{\dagger}_{j} ] = \delta_{ij}$. We use the following photon field expression~\cite{cohen:1987}
\begin{align}
\label{eq:photon}
\vec{A}(\vec{r}) = \sum_{i} \left[ \frac{\hbar}{2 \,\epsilon_{0}\, \omega_{i} L^{3}} \right]^{1/2} \left[\hat{a}_{i}\, \vec{\varepsilon}_{i} \, e^{\text{i} \vec{k}_{i} \cdot \vec{r}} + \hat{a}^{\dagger}_{i}\, \vec{\varepsilon}_{i} \, e^{-\text{i} \vec{k}_{i} \cdot \vec{r}} \right] \,.
\end{align}
We quantize the photon field $\vec{A}(\vec{r})$ in a box  of volume $L^{3}$ with a normalization condition $\vec{k} \cdot \vec{L} = 2\pi$.
Note that the wave number and the angular frequency of the photon are related via $ \omega = |\vec{k}|\, c$. 

When the photon passes through the millicharge particles, it
experiences the phase shift $\delta$. This depends on how strong the
probe photon interacts with the millicharge particles. In other
words, the relevant part of the Hamiltonian  responsible for the
phase shift is $H_{I} = H_{I1} + H_{I2}$. The first term $H_{I1}$ is
suppressed by the millicharge velocity. Moreover, it is proportional
to $(\hat{a} + \hat{a}^{\dagger})$ which induces the energy transition in a bound system. However, for a free particle system, there is no
such transition otherwise the energy conservation would be violated.
Thus, we can neglect this term for free mCPs system under consideration.

The second term $H_{I2}$ is proportional to $(\hat{a} \hat{a} + \hat{a} \hat{a}^{\dagger} + \hat{a}^{\dagger} \hat{a} + \hat{a}^{\dagger} \hat{a}^{\dagger})$ which induces two photons transition. Both of the
first and the last term violate photon number and energy
conservation for free particle system. Therefore, only the second and
the third term remain and we have
\begin{align}
\label{eq:HintF}
H_{\text{int}} &= \sum_{\alpha} \frac{q^{2}_{\alpha}}{2\,m_{\alpha}} \, \left[ \frac{\hbar}{2 \,\epsilon_{0}\, \omega_{i} L^{3}} \right] 2\left(\hat{a}^{\dagger} \hat{a} + \frac{1}{2} \right)\\
 &= \frac{\epsilon^{2} \, e^{2}}{m_{\text{Q}}}\,\left[ \frac{\hbar\,\omega^{2}}{16\,\pi^{3} \,\epsilon_{0}\, c^{3}} \right] \left(\hat{a}^{\dagger} \hat{a} + \frac{1}{2} \right)\, N_{\text{Q}}\,, 
\end{align} 
where we have assumed all mCPs have the same charge $q_{\alpha} = \epsilon\,e$ and the same mass $m_{\alpha} = m_{\text{Q}}$ such that
the sum over mCPs is proportional to the total number of mCPs $N_{Q}$. We
only
consider a single photon mode which is well approximated by
laser relevant for our proposal.
\begin{figure}
	\centering
	\includegraphics[width=0.7\textwidth]{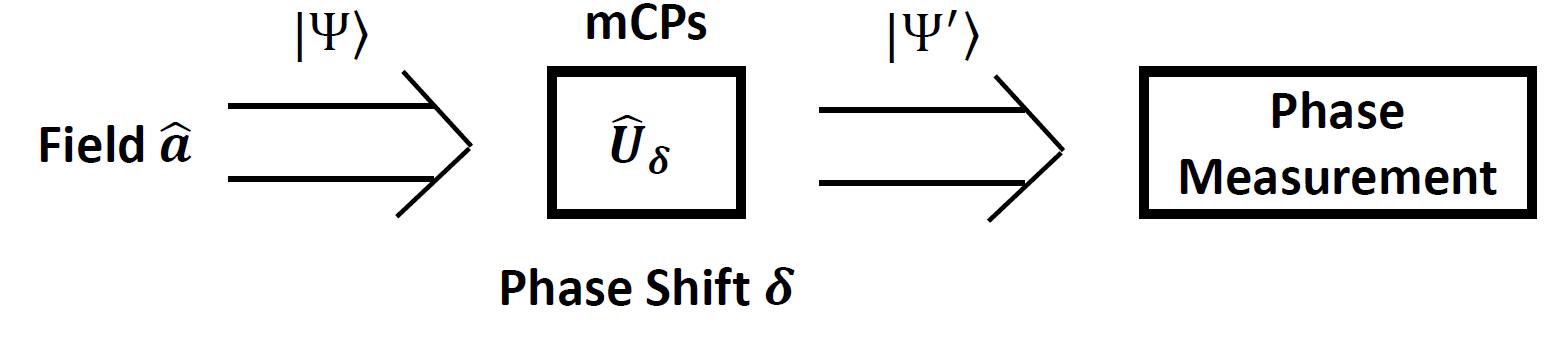}
	\caption{The phase shift $\delta$ on optical field $\hat{a}$ changes the photon state from $\ket{\Psi}$ to $\ket{\Psi^{'}}$ due to photon-mCPs interaction.}
	\label{fig:interfero}
\end{figure}
We would
like to detect the phase shift
induced by the mCPs-photon interaction using
the phase measurement scheme illustrated in
Fig.~\ref{fig:interfero}. The change of the photon state from $\ket{\Psi}$ to $\ket{\Psi^{'}}$ occurs via the unitary operator $\hat{U}_{\delta}$~\cite{Kartner:1993,Ou:2017}
\begin{align}
\label{eq:psiP}
\ket{\Psi^{'}} &= \hat{U}_{\delta} \, \ket{\Psi} =e^{\text{i}\,\hat{H}_{\text{int}}\text{t}/\hbar}\, \ket{\Psi}= e^{\text{i}\,\hat{n}\delta}\,\ket{\Psi} 
\end{align}
where $ \hat{n} \equiv \hat{a}^{\dagger}\hat{a}$ is the photon number operator which has the average value
$ n \equiv \left\langle \hat{a}^{\dagger}\hat{a} \right\rangle\gg 1$. 
 
\begin{figure}
	\centering
	\includegraphics[width=0.7\textwidth]{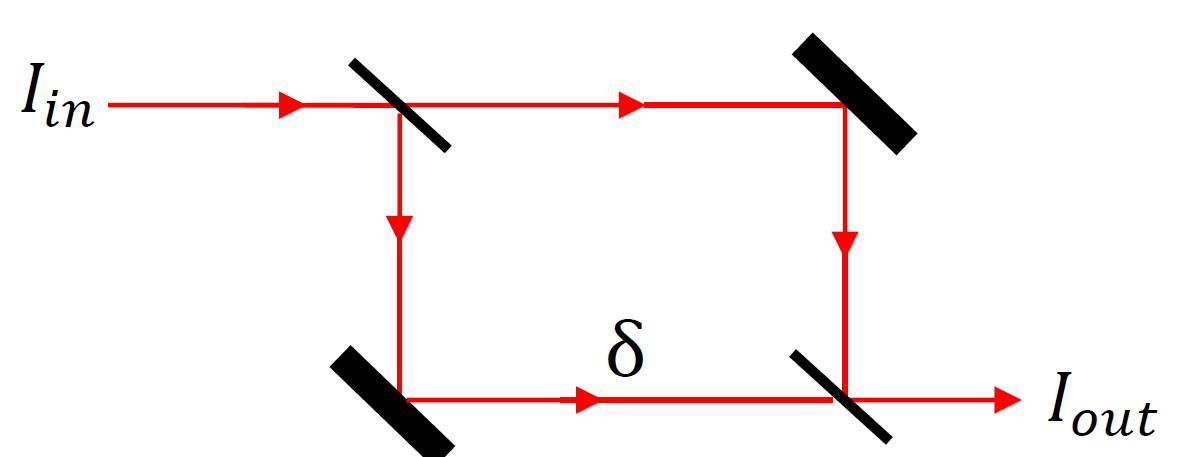}
	\caption{Mach-Zehnder interferometer with a phase shift $\delta$ in one of its arm.}
	\label{fig:heisenberg}
\end{figure}

However, due to the quantum nature of the light, there is a
limitation that prevents us to probe the phase shift as accurate as
possible. To demonstrate this, consider a simple interferometer
shown in Fig.~\ref{fig:heisenberg}. 
The interferometer is adjusted in such a way that if there is no phase shift the output intensity would be zero~\cite{Ou:2017}
\begin{align}
\label{eq:simpleInterfero}
I_{\text{out}} = I_{\text{in}}\, (1 - \text{cos}\,\delta)/2\,.
\end{align}
For a well defined input intensity $I_{\text{in}}$, the change in
the output intensity $\Delta I_{\text{out}}$ comes solely from the
phase change $\Delta \delta$
\begin{align}
\label{eq:changePhase}
\Delta I_{\text{out}} = \frac{I_{\text{in}}}{2}\, \Delta \delta\, \text{sin}\, \delta\,.
\end{align}
The sensitivity to detect this change is maximized when $\delta = \pi/2$. Since the intensity can be written in term of the photon number, one has
\begin{align}
\label{eq:changeMax}
\Delta N^{\text{max}}_{\text{out}} = \frac{N_{\text{in}}}{2}\, \Delta \delta\,,
\end{align}
where $N_{\text{in}}$ is the total photon number in the input and $\Delta N_{\text{out}}$ stands for the change in output photon number. In quantum mechanics, the lowest possible $\Delta N_{\text{out}}$ is
one. Thus, the minimum detectable phase is given by the Heisenberg
limit~\cite{Dirac:1927, Heitler:1954}
\begin{align}
\label{eq:HeisLimit}
\Delta \delta \geq \frac{1}{N}\,,
\end{align}
where $N = N_{\text{in}}/2$ is the total number of photon in one of
the interferometer arm that encounters the phase shift. Typically,
the number of photon in laser interferometer is of the
order of $10^{20}$ or larger allowing
us to detect a very tiny phase shift in the laboratory experiment.

\section{Phase Measurement Scheme}
\label{sec:phase}
\begin{figure}
	\centering
	\includegraphics[width=0.8\textwidth]{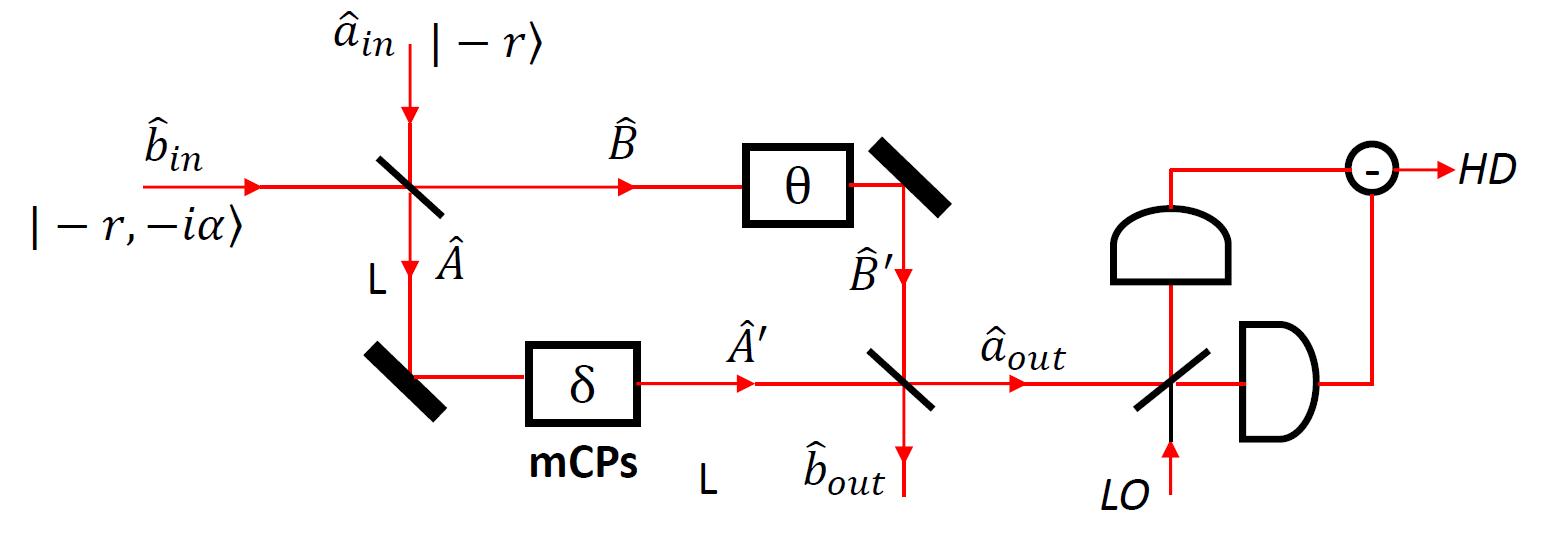}
	\caption{Mach-Zehnder interferometer used in~\cite{Bondurant:1984} with squeezed coherent state and squeezed vacuum in its input. The phase shift $\delta$ induced by mCPs-photon interaction is measured at the outpot port by homodyne detection (HD) via local oscillator (LO). }
	\label{fig:MZInterferometer}
\end{figure}

In practice, there are several interferometers that achieve the
Heisenberg limit~\cite{Bondurant:1984,Grangier:1987,Xiao:1987}. One of them is the Mach-Zehnder interferometer
shown in Fig.~\ref{fig:MZInterferometer}. We propose to use this
interferometer to measure the change of the photon phase due to its
interaction with mCPs. 

The Mach-Zehnder (MZ) interferometer under consideration have two input
ports. One of the ports is injected by a squeezed vacuum state $\ket{-r}$ while another port is fed by using a squeezed coherent
state $\ket{-r,-i\alpha}$~\cite{Bondurant:1984}, see Fig.~\ref{fig:MZInterferometer}. Here, 
the squeezed vacuum state is defined by
\begin{align}
\label{eq:defsqueeze}
\ket{r} = e^{r (\hat{a}^{\dagger\,2} - \hat{a}^{2})/2} \ket{0} \equiv \hat{S}(r) \ket{0}\,,
\end{align}
where $\hat{S}(r)$ is the squeezing operator with a real positive
value of squeezing parameter $r$. In the photon number basis $\ket{n}$,
the
coherent state $\ket{\alpha}$ can be written as
\begin{align}
\label{eq:DefCoherent}
\ket{\alpha} = e^{-|\alpha|^{2}/2} \sum^{\infty}_{n = 0} \frac{\alpha^{n}}{\sqrt{n !}}\, \ket{n} \equiv \hat{D}(\alpha) \ket{0}\,, 
\end{align} 
where $\hat{D}(\alpha) = e^{\alpha\hat{a}^{\dagger}-\alpha^{*} \hat{a}}$ is the diplacement operator. Both of the squeezing
operator as well as the displacement operator have the following
effects when acting on the
annihilation operator $\hat{a}$ and the creation operator $\hat{a}^{\dagger}$
\begin{align}
\label{eq:OpProperties}
\hat{S}^{\dagger}(r)\, \hat{a}\, \hat{S}(r) &= \hat{a}\, \text{cosh}\,r \, + \hat{a}^{\dagger} \, \text{sinh}\,r\,, \\
\hat{S}^{\dagger}(r)\, \hat{a}^{\dagger}\, \hat{S}(r) &= \hat{a}^{\dagger}\, \text{cosh}\,r \, + \hat{a} \, \text{sinh}\,r\,, \\
\hat{D}^{\dagger}(\alpha)\, \hat{a}\,\hat{D}(\alpha) &= \hat{a} \, +\, \alpha\,, \\
\hat{D}^{\dagger}(\alpha)\, \hat{a}^{\dagger}\,\hat{D}(\alpha) &= \hat{a}^{\dagger} \, +\, \alpha^{*}\,.
\end{align}
Using these operators, one can construct the squeezed coherent state  $\ket{r,\alpha}$
\begin{align}
\label{eq:SCoh}
\ket{r,\alpha} = \hat{S}(r)\,\hat{D}(\alpha)\,\ket{0}\,.
\end{align}  
In MZ interferometer, two input fields enter the first 50:50 beam splitter which
divides the photon path associated with two field operators $\hat{A}$ and $\hat{B}$. Next, the operator $\hat{A}$ transforms into $\hat{A}^{'}$  due to its interaction with mCPs while another
operator  $\hat{B}$ becomes $\hat{B}^{'}$ because of the
interferometer
adjustment. The later is set such that in the absence of mCPs-photon
interaction, the dark fringe output located at $\hat{a}_{out}$
would read the squeezed coherent state input $\ket{-r, -i\alpha}$.
The input annihilation operator $\hat{a}_{in}$ acts on the state $\ket{-r}$ while $\hat{b}_{in}$ operates on $\ket{-r,-i\alpha}$. Furthermore, the operators $\hat{A}$, $\hat{B}$, $\hat{A}^{'}$, $\hat{B}$, $\hat{a}_{out}$, and $\hat{b}_{out}$ appear in Fig.\ref{fig:MZInterferometer} are
\begin{align}
\label{eq:ABDefinition}
\hat{A} &= \frac{(\hat{a}_{in} + \hat{b}_{in})}{\sqrt{2}},\,\,\,\,\,\hat{B} = \frac{(-\hat{a}_{in} + \hat{b}_{in})}{\sqrt{2}}\,,\\
\hat{A}^{'} &= \hat{A}\,e^{\text{i}\delta},\,\,\,\,\,\hat{B}^{'} = \hat{B}\,e^{\text{i}\theta}\,,\\
\hat{a}_{out} &= \frac{(\hat{A}^{'} - \hat{B}^{'})}{\sqrt{2}},\,\,\,\,\,\hat{b}_{out} = \frac{(\hat{A}^{'} + \hat{B}^{'})}{\sqrt{2}}\,.
\end{align}
We set $\theta = \pi$ such that the dark fringe output becomes
\begin{align}
\label{eq:aout}
\hat{a}_{out} = \text{i}\, e^{i\delta/2}\, \left(\hat{a}_{in}\,\text{sin}\,\frac{\delta}{2}-\text{i}\,\hat{b}_{in}\, \text{cos}\,\frac{\delta}{2} \right)\,.
\end{align}
We measure the quadrature amplitude $\hat{X}_{a} = \hat{a}_{out} + \hat{a}^{\dagger}_{out}$ at the output by homodyne detection (HD) and have
\begin{align}
\label{eq:XaDelta}
\hat{X}_{a} = -\text{sin}\,\frac{\delta}{2}\,\hat{Y}_{a_{in}}(-\delta/2) + \text{cos}\,\frac{\delta}{2}\,\hat{X}_{b_{in}}(-\delta/2)\,.
\end{align}
Here, we have defined $\hat{Y}_{a_{in}}(-\delta/2)$ and $\hat{X}_{b_{in}}(-\delta/2)$ as
\begin{align}
\label{eq:XY}
\hat{Y}_{a_{in}}(-\delta/2) &= -\text{i}\,\left(\hat{a}_{in}\,e^{\text{i}\,\delta/2} - \hat{a}^{\dagger}_{in}\,e^{-\text{i}\,\delta/2} \right)\,,\\
\hat{X}_{b_{in}}(-\delta/2) &= \left(\hat{b}_{in}\,e^{\text{i}\,\delta/2} + \hat{b}^{\dagger}_{in}\,e^{-\text{i}\,\delta/2} \right)\,.
\end{align}
To determine the signal to noise ratio (SNR) of our phase
measurement, one needs the expectation value of the quadrature
amplitude $\hat{X}_{a}$ as well as its fluctuation $\Delta^{2} \hat{X}_{a}$
\begin{align}
\label{eq:evXa}
\left\langle\hat{X}_{a}\right\rangle &= \alpha\,(\mu + \nu)\,\text{sin}\,\delta\,\\
\left\langle\Delta^{2} \hat{X}_{a}\right\rangle &= (\mu - \nu)^{2}\,,
\end{align}
with $\mu = \text{cosh}\,r$ and $\nu = \text{sinh}\,r$. The signal to noise ratio (SNR) for the observable $\hat{X}_{a}$ reads~\cite{Ou:2017}
\begin{align}
\label{eq:SNR}
\text{SNR} \equiv \frac{\left\langle\hat{X}_{a}\right\rangle^{2}}{\left\langle\Delta^{2} \hat{X}_{a}\right\rangle} = \frac{\alpha^{2}\,(\mu + \nu)^{2}\, \text{sin}^{2}\, \delta}{(\mu - \nu)^{2}}\,.
\end{align}
The phase sensing photon number is defined as $N_{ps} \equiv \left\langle\hat{A}^{\dagger} \hat{A}\right\rangle = \nu^{2} + \alpha^{2} (\mu + \nu)^{2} /2$. We keep $N_{ps}$ constant and take $\nu \gg 1$ such that $\mu - \nu = 1/(\mu + \nu) \approx 1/2\nu$. Thus, the SNR becomes
\begin{align}
\label{eq:SNRfin}
\text{SNR} = 8\, (N_{ps} -\nu^{2})\, \nu^{2}\,\text{sin}^{2}\,\delta \leq 2\, N^{2}_{ps}\,\text{sin}^{2}\,\delta\,,
\end{align}
reaching the maximum value when $\nu^{2} = N_{ps}/2$. In the limit $\delta \ll 1$ and $\text{SNR} \sim 1$, the minimum detectable phase shift is $\delta_{min} \sim 1/N_{ps}$ or the Heisenberg limit. From here on, we take the maximum SNR value in Eq.~\eqref{eq:SNRfin}.

\section{Results and Discussion}
\label{sec:result} 
To claim a discovery, the required SNR value must be larger than
one. We take three benchmark values of millicharge mCPs density $n_{\text{Q}}$: $1\,\text{cm}^{-3}$, $10^{3}\,\text{cm}^{-3}$, and $10^{6}\,\text{cm}^{-3}$ to compare with the projected sensitivities  set by ion trap proposal~\cite{Budker:2021quh}. Looking at Eq~\eqref{eq:HintF} and \eqref{eq:psiP}, the
phase shift is given by
\begin{align}
\label{eq:delta}
\delta = \frac{\epsilon^{2} \, e^{2}}{m_{\text{Q}}}\,\left[ \frac{\omega^{2}}{16\,\pi^{3} \,\epsilon_{0}\, c^{3}} \right] \, N_{\text{Q}}\,t\,.
\end{align}     
\begin{figure}
\centering
\includegraphics[width=6.9cm]{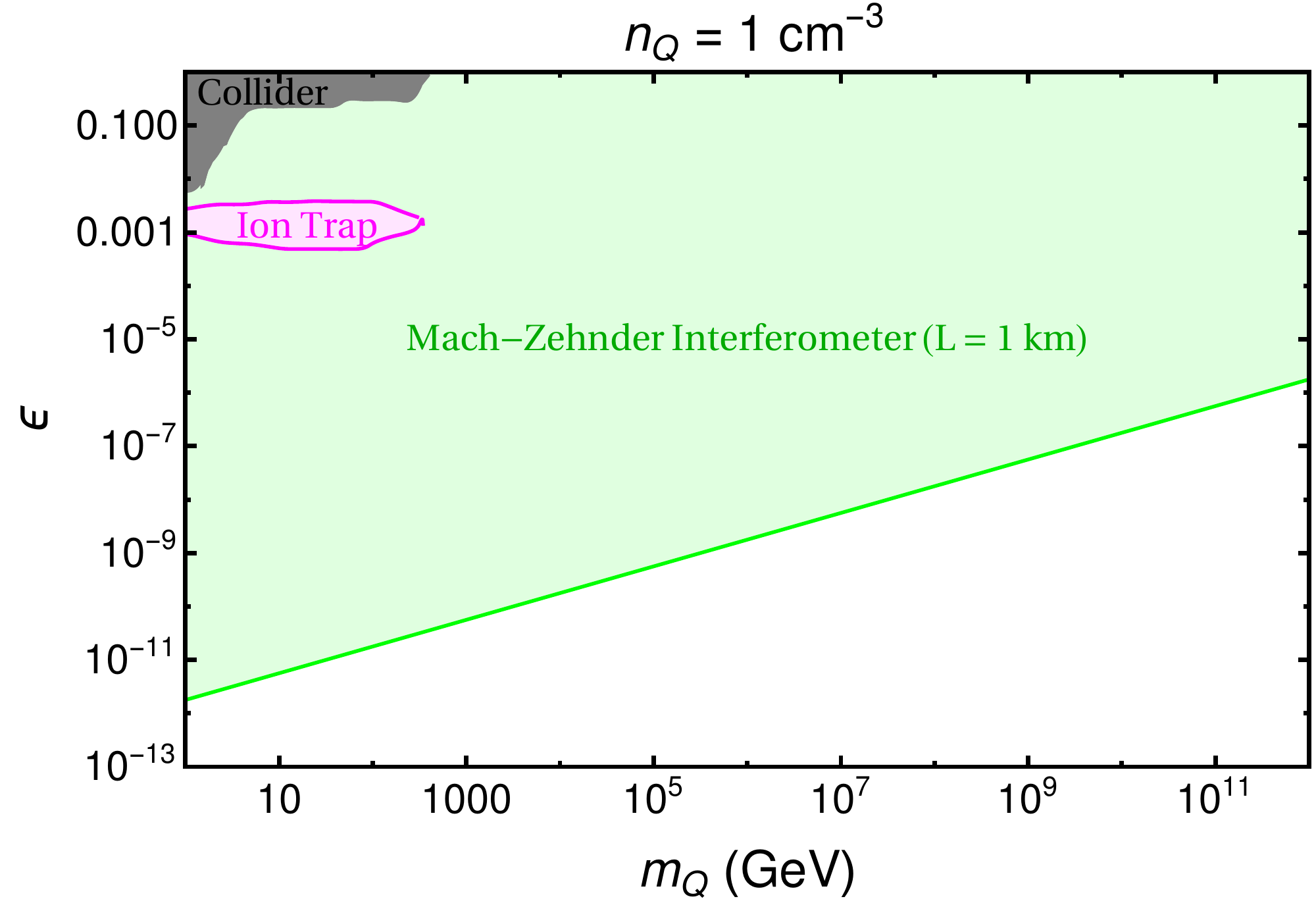}
\includegraphics[width=6.9cm]{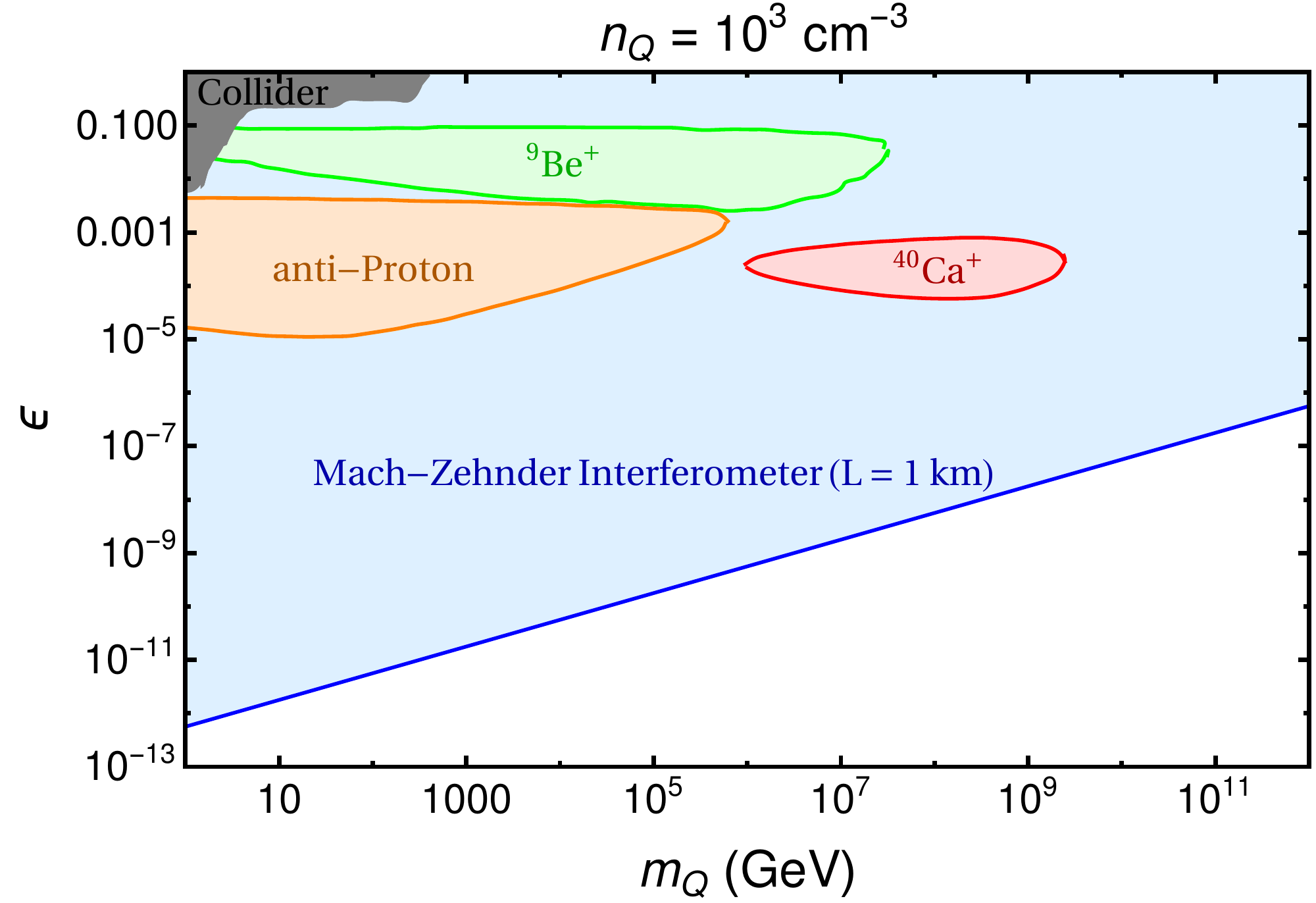}
\includegraphics[width=6.9cm]{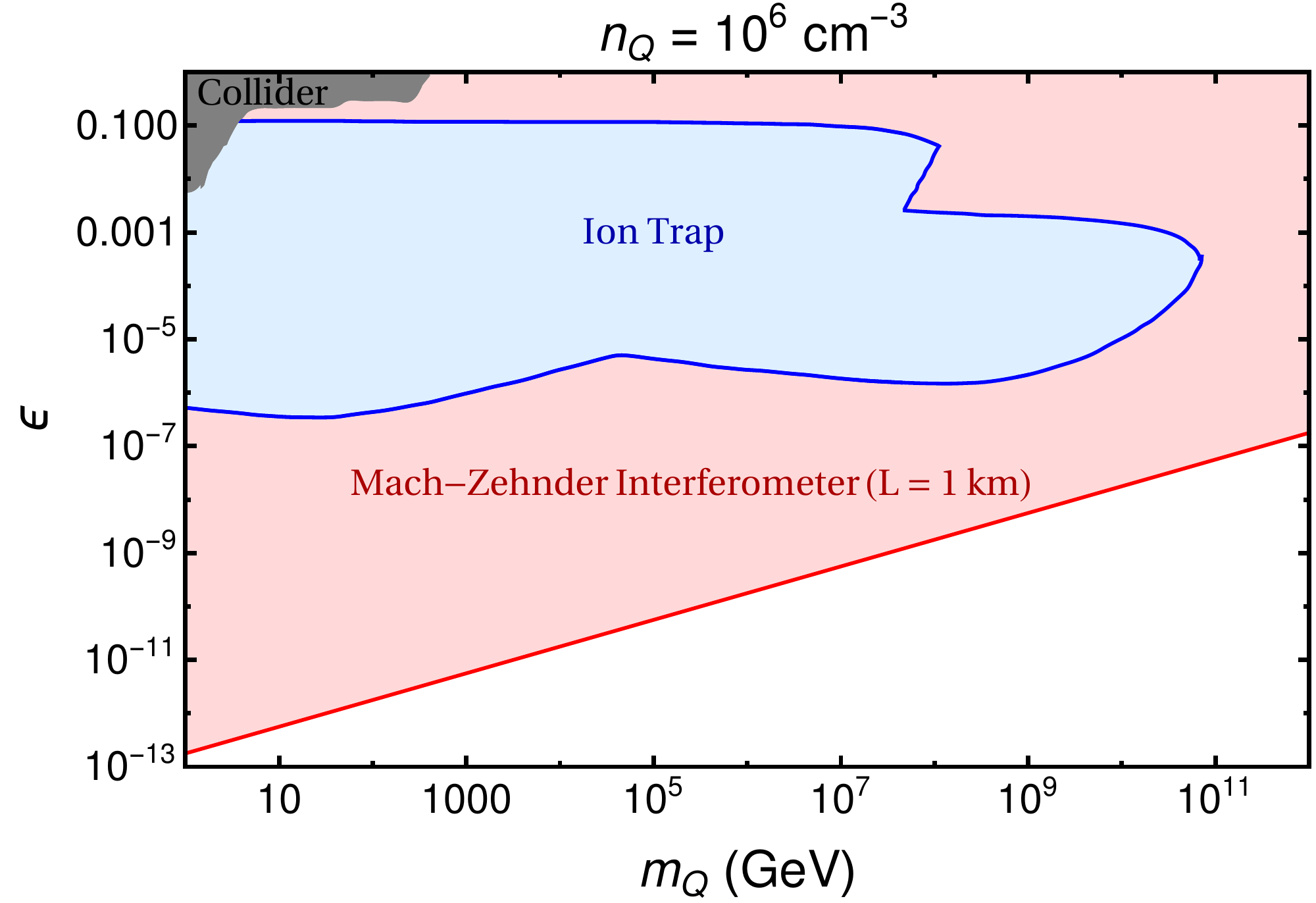}
\caption{The projected sensitivity of MZ interferometer with arm length L = 1 km and 1.17 eV laser for different mCPs densities $n_{\text{Q}}$: $1\,\text{cm}^{-3}$ (top-left), $10^{3}\,\text{cm}^{-3}$ (top-right), and $10^{6}\,\text{cm}^{-3}$ (bottom). We take the phase sensing photon number $N_{ps} = 10^{23}$~\cite{GammeVT-969:2007pci,Bahre:2013ywa,ALPS:2009des,Inada:2013tx}.} 
\label{fig:sensitivity}
\end{figure} 
Since the probe photon only interacts with mCPs along its path $\ell$, the
total mCPs number $N_{\text{Q}}$ can be obtained by integrating the
number of mCPs per unit length with respect to the total length
traversed by the photon
\begin{align}
\label{eq:ell}
N_{\text{Q}} = \int^{\text{L}}_{0} d\ell\, \tilde{n}_{\text{Q}}\,,
\end{align}
where L is the interferometer arm length and $\tilde{n}_{\text{Q}} = n^{1/3}_{\text{Q}}$ denotes the number of mCPs per unit length in $\text{cm}^{-1}$.
In this case, the number of mCPs per unit
length for three different mCPs densities considered here are $1\,\text{cm}^{-1}$, $10\,\text{cm}^{-1}$, and $100\,\text{cm}^{-1}$, respectively. The
time parameter $t$ in Eq.~\eqref{eq:delta} describes the mCPs-photon
interaction
time which is
simply $\text{L}/c$. 

The sensitivity of MZ interferometer is shown in Fig.~\ref{fig:sensitivity}. For $n_{\text{Q}} = 1\,\text{cm}^{-3}$ (the upper left panel), the
excluded region from collider search is given by the gray area in the upper-left corner.
Moreover, the parameter regime of projected  sensitivity from ion trap proposal is
shown by the pink region. The light green region corresponds to the
SNR $>$ 1 of our phase measurement scheme. We see that our proposal
is several order of magnitude more sensitive compared to the
existing limits. For higher mCPs number density, ion trap results cover several region of
parameter space in case of  $n_{\text{Q}} = 10^{3}\,\text{cm}^{-3}$
depending on the ion employed in the trap (see the upper-right panel of Fig.~\ref{fig:sensitivity}). As the number of mCPs
density gets higher, the ion trap proposal covers large area in the parameter space of coupling $\epsilon$ and mCPs mass $m_{\text{Q}}$. 
Still, for these two cases, our proposal is taking a lead on
sensitivity as can be seen from light blue area and light red area
for mCPs density equals to $10^{3}\,\text{cm}^{-3}$ and $10^{6}\,\text{cm}^{-3}$ (the upper-right and  lower panel), respectively. This shows that the MZ interferometer
can be utilized
as mCPs detector in higher mCPs mass, especially the earth
bound mCPs. 
    
\section{Summary}
\label{sec:Summary}

The existence of mCPs can be naturally realized if the dark sector communicate with the SM via a $U(1)$ mixing. It is possible that the mCPs constitute part of the dark matter and are stoped and accumulated inside the earth. As a result, the number density of mCPs inside the earth can be several orders of magnitude higher than that of the local DM. In this paper, we propose that the mCPs bound in earth can be probed by using the Mach-Zehnder interferometer (MZ interferometer) though a phase shift of laser beam when photons interact with mCPs, and focus on the case of heavy mCPs of mass $m_Q > 1$ GeV.  Notice that, for the mass of mCPs to be larger than $1$ GeV, the number density underground will be much higher than that on the surface of the earth due to gravitation pulling and  traffic jam effects. Hence, one arm of the MZ interferometer should be implemented underground (e.g. the lower horizontal beam with $\hat{A}'$ in Fig.~\ref{fig:MZInterferometer}) while the other arm on the earth surface (e.g. the upper horizontal beam with $\hat{B}'$ in Fig.~\ref{fig:MZInterferometer}). The (significant) difference of the number densities in the locations of two arms of MZ interferometer can be measured. 

Given the number density of mCPs underground, we estimate the signal to noise ratio for the phase shift, and found that the  $U(1)$ mixing parameter $\epsilon$ can be probed as low as the order of $10^{-13}$  if the number density underground is about $10^{6}~\rm cm^{-3}$ for mass of mCPs around $1$ GeV. As compared with the current bound by LHC, which is of order $10^{-3}\sim10^{-2}$, the sensitivity of our proposal is about $10^{10}$ higher for the mass of mCPs $m_Q$ up to $100$ GeV. One should also notice that for $m_Q \gtrsim 1$ GeV, astrophysical observations and beam-dump experiments have no sensitivity. Even compared with the novel detection approach using ion trap recently proposed by~\cite{Budker:2021quh}, MZ interferometer can cover whole parameter space probed by ion trap experiments.


\section*{Acknowledgment}
\label{sec:Acknowledgment}
We would like to acknowledge the support of National Center for Theoretical Sciences (NCTS). This work was supported in part by the National Science and Technology Council (NSTC) of Taiwan under Grant No.MOST 110-2112-M-003-003-, 111-2112-M-003-006 and 111-2811-M-003-025-.


\begin{thebibliography}{99}

\bibitem{Dvorkin:2019zdi}
C.~Dvorkin, T.~Lin and K.~Schutz,
Phys. Rev. D \textbf{99}, no.11, 115009 (2019)
[erratum: Phys. Rev. D \textbf{105}, no.11, 119901 (2022)]
doi:10.1103/PhysRevD.99.115009
[arXiv:1902.08623 [hep-ph]].

\bibitem{Creque-Sarbinowski:2019mcm}
C.~Creque-Sarbinowski, L.~Ji, E.~D.~Kovetz and M.~Kamionkowski,
Phys. Rev. D \textbf{100}, no.2, 023528 (2019)
doi:10.1103/PhysRevD.100.023528
[arXiv:1903.09154 [astro-ph.CO]].


\bibitem{Prinz:1998ua}
A.~A.~Prinz, R.~Baggs, J.~Ballam, S.~Ecklund, C.~Fertig, J.~A.~Jaros, K.~Kase, A.~Kulikov, W.~G.~J.~Langeveld and R.~Leonard, \textit{et al.}
Phys. Rev. Lett. \textbf{81}, 1175-1178 (1998)
doi:10.1103/PhysRevLett.81.1175
[arXiv:hep-ex/9804008 [hep-ex]].

\bibitem{Magill:2018tbb}
G.~Magill, R.~Plestid, M.~Pospelov and Y.~D.~Tsai,
Phys. Rev. Lett. \textbf{122}, no.7, 071801 (2019)
doi:10.1103/PhysRevLett.122.071801
[arXiv:1806.03310 [hep-ph]].

\bibitem{Marocco:2020dqu}
G.~Marocco and S.~Sarkar,
SciPost Phys. \textbf{10}, no.2, 043 (2021)
doi:10.21468/SciPostPhys.10.2.043
[arXiv:2011.08153 [hep-ph]].


\bibitem{Ball:2020dnx}
A.~Ball, G.~Beauregard, J.~Brooke, C.~Campagnari, M.~Carrigan, M.~Citron, J.~De La Haye, A.~De Roeck, Y.~Elskens and R.~E.~Franco, \textit{et al.}
Phys. Rev. D \textbf{102}, no.3, 032002 (2020)
doi:10.1103/PhysRevD.102.032002
[arXiv:2005.06518 [hep-ex]].

\bibitem{ArgoNeuT:2019ckq}
R.~Acciarri \textit{et al.} [ArgoNeuT],
Phys. Rev. Lett. \textbf{124}, no.13, 131801 (2020)
doi:10.1103/PhysRevLett.124.131801
[arXiv:1911.07996 [hep-ex]].

\bibitem{Davidson:2000hf}
S.~Davidson, S.~Hannestad and G.~Raffelt,
JHEP \textbf{05}, 003 (2000)
doi:10.1088/1126-6708/2000/05/003
[arXiv:hep-ph/0001179 [hep-ph]].

\bibitem{Chang:2018rso}
J.~H.~Chang, R.~Essig and S.~D.~McDermott,
JHEP \textbf{09}, 051 (2018)
doi:10.1007/JHEP09(2018)051
[arXiv:1803.00993 [hep-ph]].
\bibitem{Knapen:2017ekk}
S.~Knapen, T.~Lin, M.~Pyle and K.~M.~Zurek,
Phys. Lett. B \textbf{785}, 386-390 (2018)
doi:10.1016/j.physletb.2018.08.064
[arXiv:1712.06598 [hep-ph]].

\bibitem{Blanco:2019lrf}
C.~Blanco, J.~I.~Collar, Y.~Kahn and B.~Lillard,
Phys. Rev. D \textbf{101}, no.5, 056001 (2020)
doi:10.1103/PhysRevD.101.056001
[arXiv:1912.02822 [hep-ph]].

\bibitem{Essig:2019kfe}
R.~Essig, J.~P\'erez-R\'\i{}os, H.~Ramani and O.~Slone,
Phys. Rev. Research. \textbf{1}, 033105 (2019)
doi:10.1103/PhysRevResearch.1.033105
[arXiv:1907.07682 [hep-ph]].

\bibitem{Berlin:2019uco}
A.~Berlin, R.~T.~D'Agnolo, S.~A.~R.~Ellis, P.~Schuster and N.~Toro,
Phys. Rev. Lett. \textbf{124}, no.1, 011801 (2020)
doi:10.1103/PhysRevLett.124.011801
[arXiv:1908.06982 [hep-ph]].

\bibitem{Kurinsky:2019pgb}
N.~A.~Kurinsky, T.~C.~Yu, Y.~Hochberg and B.~Cabrera,
Phys. Rev. D \textbf{99}, no.12, 123005 (2019)
doi:10.1103/PhysRevD.99.123005
[arXiv:1901.07569 [hep-ex]].

\bibitem{SENSEI:2020dpa}
L.~Barak \textit{et al.} [SENSEI],
Phys. Rev. Lett. \textbf{125}, no.17, 171802 (2020)
doi:10.1103/PhysRevLett.125.171802
[arXiv:2004.11378 [astro-ph.CO]].

\bibitem{Griffin:2020lgd}
S.~M.~Griffin, Y.~Hochberg, K.~Inzani, N.~Kurinsky, T.~Lin and T.~Chin,
Phys. Rev. D \textbf{103}, no.7, 075002 (2021)
doi:10.1103/PhysRevD.103.075002
[arXiv:2008.08560 [hep-ph]].


\bibitem{XENON:2020rca}
E.~Aprile \textit{et al.} [XENON],
Phys. Rev. D \textbf{102}, no.7, 072004 (2020)
doi:10.1103/PhysRevD.102.072004
[arXiv:2006.09721 [hep-ex]].
\bibitem{Neufeld:2018slx}
D.~A.~Neufeld, G.~R.~Farrar and C.~F.~McKee,
Astrophys. J. \textbf{866}, no.2, 111 (2018)
doi:10.3847/1538-4357/aad6a4
[arXiv:1805.08794 [astro-ph.CO]].

\bibitem{Pospelov:2020ktu}
M.~Pospelov and H.~Ramani,
Phys. Rev. D \textbf{103}, no.11, 115031 (2021)
doi:10.1103/PhysRevD.103.115031
[arXiv:2012.03957 [hep-ph]].

\bibitem{Tsuchida:2019hhc}
S.~Tsuchida, N.~Kanda, Y.~Itoh and M.~Mori,
Phys. Rev. D \textbf{101} (2020) no.2, 023005
[arXiv:1909.00654 [astro-ph.HE]].

\bibitem{Lee:2020dcd}
C.~H.~Lee, C.~S.~Nugroho and M.~Spinrath,
Eur. Phys. J. C \textbf{80}, no.12, 1125 (2020)
doi:10.1140/epjc/s10052-020-08692-3
[arXiv:2007.07908 [hep-ph]].

\bibitem{Chen:2021apc}
C.~R.~Chen and C.~S.~Nugroho,
Phys. Rev. D \textbf{105}, no.8, 083001 (2022)
doi:10.1103/PhysRevD.105.083001
[arXiv:2111.11014 [hep-ph]].

\bibitem{Ismail:2022ukp}
M.~A.~Ismail, C.~S.~Nugroho and H.~T.~K.~Wong,
[arXiv:2211.13384 [hep-ph]].

\bibitem{Lee:2022tsw}
C.~H.~Lee, R.~Primulando and M.~Spinrath,
[arXiv:2208.06232 [hep-ph]].

\bibitem{Seto:2004zu}
N.~Seto and A.~Cooray,
Phys. Rev. D \textbf{70} (2004), 063512
[arXiv:astro-ph/0405216 [astro-ph]].

\bibitem{Adams:2004pk}
A.~W.~Adams and J.~S.~Bloom,
[arXiv:astro-ph/0405266 [astro-ph]].

\bibitem{Riedel:2012ur}
C.~J.~Riedel,
Phys.\ Rev.\ D {\bf 88} (2013) no.11,  116005
[arXiv:1212.3061 [quant-ph]].

\bibitem{PhysRevLett.114.161301}
Y.~V.~Stadnik and V.~V.~Flambaum,
Phys.\ Rev.\ Lett.\  {\bf 114} (2015) 161301
[arXiv:1412.7801 [hep-ph]].

\bibitem{Arvanitaki:2015iga}
A.~Arvanitaki, S.~Dimopoulos and K.~Van Tilburg,
Phys.\ Rev.\ Lett.\  {\bf 116} (2016) no.3,  031102
[arXiv:1508.01798 [hep-ph]].

\bibitem{Stadnik:2015xbn}
Y.~V.~Stadnik and V.~V.~Flambaum,
Phys.\ Rev.\ A {\bf 93} (2016) no.6,  063630
[arXiv:1511.00447 [physics.atom-ph]].

\bibitem{Branca:2016rez}
A.~Branca {\it et al.},
Phys.\ Rev.\ Lett.\  {\bf 118} (2017) no.2,  021302
[arXiv:1607.07327 [hep-ex]].

\bibitem{Riedel:2016acj}
C.~J.~Riedel and I.~Yavin,
Phys.\ Rev.\ D {\bf 96} (2017) no.2,  023007
[arXiv:1609.04145 [quant-ph]].

\bibitem{Hall:2016usm}
E.~D.~Hall, R.~X.~Adhikari, V.~V.~Frolov, H.~Müller, and M.~Pospelov,
Phys. Rev. D \textbf{98} (2018) no.8, 083019
[arXiv:1605.01103 [gr-qc]].

\bibitem{Jung:2017flg}
S.~Jung and C.~S.~Shin,
Phys.\ Rev.\ Lett.\  {\bf 122} (2019) no.4,  041103
[arXiv:1712.01396 [astro-ph.CO]].

\bibitem{Pierce:2018xmy}
A.~Pierce, K.~Riles and Y.~Zhao,
Phys.\ Rev.\ Lett.\  {\bf 121} (2018) no.6,  061102
[arXiv:1801.10161 [hep-ph]].

\bibitem{Morisaki:2018htj}
S.~Morisaki and T.~Suyama,
Phys.\ Rev.\ D {\bf 100} (2019) no.12,  123512
[arXiv:1811.05003 [hep-ph]].

\bibitem{Grote:2019uvn}
H.~Grote and Y.~V.~Stadnik,
Phys.\ Rev.\ Research.\  {\bf 1} (2019) 033187
[arXiv:1906.06193 [astro-ph.IM]].

\bibitem{Okun:1982xi}
L.~B.~Okun,
Sov. Phys. JETP \textbf{56}, 502 (1982)
ITEP-48-1982;
B.~Holdom,
Phys. Lett. B \textbf{166}, 196-198 (1986)
doi:10.1016/0370-2693(86)91377-8;
E.~Izaguirre and I.~Yavin,
Phys. Rev. D \textbf{92}, no.3, 035014 (2015)
doi:10.1103/PhysRevD.92.035014
[arXiv:1506.04760 [hep-ph]].

\bibitem{Budker:2021quh}
D.~Budker, P.~W.~Graham, H.~Ramani, F.~Schmidt-Kaler, C.~Smorra and S.~Ulmer,
PRX Quantum \textbf{3}, no.1, 010330 (2022)
doi:10.1103/PRXQuantum.3.010330
[arXiv:2108.05283 [hep-ph]].

\bibitem{cohen:1987}
C.~C.~Tannoudji, J.~D.~Roc, and G.~Grynberg,
Photons and Atoms Introduction to Quantum Electrodynamics,
John Wiley and Sons (1987),
doi:https://doi.org/10.1002/9783527618422.ch3

\bibitem{Kartner:1993}
F.~X.~Kartner and H.~A.~Haus,
Phys.~Rev.~A \textbf{65}, p. 4585.

\bibitem{Ou:2017}
Z.~Y.~J.~Ou,
Quantum Optics for Experimentalists,
World Scientific (2017)

\bibitem{Dirac:1927}
P.~A.~M.~Dirac
Proc. R. Soc. London Ser. A \textbf{114}, p.243.

\bibitem{Heitler:1954}
W.~Heitler, The Quantum Theory of Radiation, 3rd edn., Oxford University Press, London. 

\bibitem{Bondurant:1984}
R.~S.~Bondurant and J.~H.~Shapiro
Phys.~Rev.~D \textbf{30}, p. 2548.

\bibitem{Grangier:1987}
P.~Grangier, R.~E.~Slusher, B.~Yurke, and A.~Laporta,
Phys. Rev. Lett. \textbf{59}, p.2153

\bibitem{Xiao:1987}
M.~Xiao, L.~A.~Wu, and H.~J.~Kimble
Phys. Rev. Lett. \textbf{59}, p.278

\bibitem{GammeVT-969:2007pci}
A.~S.~Chou \textit{et al.} [GammeV (T-969)],
Phys. Rev. Lett. \textbf{100}, 080402 (2008)
doi:10.1103/PhysRevLett.100.080402
[arXiv:0710.3783 [hep-ex]].

\bibitem{Bahre:2013ywa}
R.~B\"ahre, B.~D\"obrich, J.~Dreyling-Eschweiler, S.~Ghazaryan, R.~Hodajerdi, D.~Horns, F.~Januschek, E.~A.~Knabbe, A.~Lindner and D.~Notz, \textit{et al.}
JINST \textbf{8}, T09001 (2013)
doi:10.1088/1748-0221/8/09/T09001
[arXiv:1302.5647 [physics.ins-det]].

\bibitem{ALPS:2009des}
K.~Ehret \textit{et al.} [ALPS],
Nucl. Instrum. Meth. A \textbf{612}, 83-96 (2009)
doi:10.1016/j.nima.2009.10.102
[arXiv:0905.4159 [physics.ins-det]].

\bibitem{Inada:2013tx}
T.~Inada, T.~Namba, S.~Asai, T.~Kobayashi, Y.~Tanaka, K.~Tamasaku, K.~Sawada and T.~Ishikawa,
Phys. Lett. B \textbf{722}, 301-304 (2013)
doi:10.1016/j.physletb.2013.04.033
[arXiv:1301.6557 [physics.ins-det]].






\end{thebibliography}
\end{document}